\documentclass[12pt]{article}
\usepackage{graphicx}
\usepackage{multirow}
\usepackage{xspace}
\usepackage{lineno}
\usepackage{subcaption}

\newcommand\pubnumber{Proceedings of Top2017 \\ ATL-PHYS-PROC-2017-259}
\newcommand\pubdate{\today}

\def\institute{Max-Planck-Institute for Physics\\
D-80805 Munich, GERMANY}

\def\Title#1{\begin{center} {\Large #1 } \end{center}}
\def\Author#1{\begin{center}{ \sc #1} \end{center}}
\def\Address#1{\begin{center}{ \it #1} \end{center}}

\newcommand\pubblock{\rightline{\begin{tabular}{l} \pubnumber\\
         \pubdate  \end{tabular}}}
\newenvironment{Abstract}{\begin{quotation}  }{\end{quotation}}
\newenvironment{Presented}{\begin{quotation} \begin{center} 
             PRESENTED AT\end{center}\bigskip 
      \begin{center}\begin{large}}{\end{large}\end{center} \end{quotation}}

\newcommand{\TeV}{\ensuremath{\mathrm{TeV}}}
\newcommand{\GeV}{\ensuremath{\mathrm{GeV}}}
\newcommand{\XZ           }[3]{\ensuremath{#1\,\pm\, #2~\mathrm{(stat)}\,\pm\, #3~\mathrm{(syst)}}\xspace}
\newcommand{\XZtot        }[2]{\ensuremath{#1\,\pm\, #2}\xspace}
\newcommand{\ptlb         }{\ensuremath{p_{T,\ell b}}\xspace}
\newcommand{\mtop           }{\ensuremath{m_{\mathrm{top}}}\xspace}
\newcommand{\mtpole           }{\ensuremath{m_{\mathrm{top}}^{\mathrm{pole}}}\xspace}
\newcommand{\mlb          }{\ensuremath{m_{\ell b}}\xspace}
\newcommand{\mtr          }{\ensuremath{\mtop^{\mathrm{reco}}}\xspace}
\newcommand{\mWr          }{\ensuremath{m_{W}^{\mathrm{reco}}}\xspace}
\newcommand{\rbqr         }{\ensuremath{R_{bq}^{\mathrm{reco}}}\xspace}
\newcommand{\CombVal                   }{\ensuremath{172.51}\xspace}
\newcommand{\CombSta                   }{\ensuremath{0.27}\xspace}
\newcommand{\CombSys                   }{\ensuremath{0.42}\xspace}
\newcommand{\CombUnc                   }{\ensuremath{0.50}\xspace}
\newcommand{\CombUncStab               }{\ensuremath{0.04}\xspace}
\newcommand{\antibar}[1]{\ensuremath{#1\bar{#1}}\xspace}
\newcommand{\ttbar}{\antibar{t}}
\newcommand{\ttbarjj      }{\ensuremath{\ttbar\to\textrm{all-jets}}\xspace}
\newcommand{\ttbarll      }{\ensuremath{\ttbar\to\textrm{dilepton}}\xspace}
\newcommand{\ttbarlj      }{\ensuremath{\ttbar\to\textrm{lepton+jets}}\xspace}

\def\beq{\begin{equation}}
\def\eeq#1{\label{#1}\end{equation}}
\def\eeqn{\end{equation}}

\def\beqa{\begin{eqnarray}}
\def\eeqa#1{\label{#1}\end{eqnarray}}
\def\eeqan{\end{eqnarray}}

\let\bar=\overbar

\def\Dslash{\not{\hbox{\kern-4pt $D$}}}
\def\dslash{\not{\hbox{\kern-2pt $\del$}}}

\def\msb{{\bar{\ssstyle M \kern -1pt S}}}

\begin{document}
\begin{titlepage}
\pubblock

\vfill
\Title{Top quark mass in ATLAS}
\vfill
\Author{Benjamin Pearson\\
on behalf of the ATLAS Collaboration}
\Address{\institute}
\vfill
\begin{Abstract}
ATLAS has made several measurements of the top quark mass using proton-proton collision data 
recorded in 2012 at the LHC with a centre-of-mass energy of 8~\TeV. Those summarised here include an indirect determination of the top quark pole mass from lepton differential cross-sections; previous direct measurements of the top quark mass in the \ttbarll and \ttbarjj decay channels as well as in the $t$-channel of single-top-quark production; and lastly, the new direct measurement of the top quark mass in the \ttbarlj decay channel and its combination with previous measurements.
\end{Abstract}
\vfill
\begin{Presented}
$10^{th}$ International Workshop on Top Quark Physics\\
Braga, Portugal,  September 17--22, 2017
\end{Presented}
\vfill
\end{titlepage}
\def\thefootnote{\fnsymbol{footnote}}
\setcounter{footnote}{0}
%

\section{Introduction}

The top quark mass (\mtop) is an important parameter in the Standard Model (SM) and, as such, its precise measurement is helpful in evaluating the internal consistency of the SM and in testing possible extensions. At the CERN Large Hadron Collider (LHC), top quarks are primarily produced as quark-antiquark pairs (\ttbar) and in the SM the top quark decays almost exclusively to a $W$ boson and a $b$ quark. Thus, the \ttbar decays are governed by the decay modes of the $W$ bosons, namely leptonic decay $W \rightarrow \ell\nu$ to a charged lepton ($\ell$) and a neutrino ($\nu$), and hadronic decay $W \rightarrow qq$ to a pair of quarks ($q$) leading to jets. Therefore, the final state topology options for \ttbar production are denoted by \ttbarjj, \ttbarlj, and \ttbarll.

The ATLAS experiment~\cite{atlas} has made several measurements of the top quark mass using LHC proton-proton collisions at a centre-of-mass (CM) energy of 8~\TeV. This proceeding first gives a summary of an indirect determination of the top quark pole mass (\mtpole), performed using lepton differential cross-sections~\cite{lepdiff}. Second, a brief review is given of previous direct measurements of \mtop in the \ttbarll~\cite{dl8} and \ttbarjj~\cite{aj8} decay channels as well as in the $t$-channel of single-top-quark production~\cite{stop}. Lastly, a summary is given for the new direct measurement in the \ttbarlj channel, along with its combination with previous measurements~\cite{lj8}.

\section{Indirect determination of \boldmath{\mtpole}}

Using $\ttbar \rightarrow e\mu + X$ events, eight differential cross-sections are measured: the single lepton transverse momentum ($p_{\mathrm{T}}^{\ell}$) and absolute pseudorapidity ($|\eta|^{\ell}$), the dilepton system transverse momentum, invariant mass and absolute rapidity ($p_{\mathrm{T}}^{e\mu}$, $m^{e\mu}$, $|y^{e\mu}|$), the azimuthal angle between the two leptons ($\Delta{\phi^{e\mu}}$), and the scalar sum of the transverse momentum and sum of the energies of the two leptons ($p_{\mathrm{T}}^{e} + p_{\mathrm{T}}^{\mu}$ and $E^{e} + E^{\mu}$). 

Many of the measured differential cross-sections are sensitive to the top quark mass, which is extracted from the normalised distributions, after correcting for acceptance and resolution effects of the detector, by fits to QCD predictions from the MCFM program~\cite{mcfm} at next-to-leading order (NLO) in the strong coupling $\alpha_{\mathrm{s}}$. This indirect determination therefore corresponds to the top quark pole mass~\cite{polemass}. A combined fit to all of the distributions results in $\mtpole = 173.2 \pm 0.9\,\mathrm{(stat)} \pm 0.8\,\mathrm{(syst)} \pm 1.2\,\mathrm{(theo)}\,\GeV$, with a total uncertainty of 1.6~\GeV~\cite{lepdiff}. The theory uncertainty is dominated by variations in the QCD factorisation and renormalisation scales.

\section{Previous direct measurements of \boldmath{\mtop} at 8~\boldmath{\TeV}}

The previous direct measurements discussed here include the \ttbarjj~\cite{aj8}, \ttbarll~\cite{dl8}, and single-top~\cite{stop} analyses performed at a CM energy of 8~\TeV. For comparison, their results are all given in Table~\ref{tab:measurements} with a partial breakdown of their uncertainties. These direct measurements all use the template method, in which simulated distributions of variables that are sensitive to \mtop are fit to analytical functions to interpolate between several discrete values of the input \mtop. A fit to the distribution in data yields the value of \mtop that best describes the data. This mass corresponds to the mass definition used in the MC event simulation. How this relates to the top quark pole mass is an object of theoretical investigation~\cite{polemass}.

The measurement in the $t$-channel of single-top-quark production uses the invariant mass of the lepton and $b$-jet system (\mlb) as its estimator for \mtop and uses a neural network to better distinguish between signal and background events. A binned maximum-likelihood fit to the data gives $\mtop = \XZ{172.2}{0.7}{2.0}~\GeV$, where by far the largest contribution to the uncertainty comes from variations in the jet energy scale (JES), as shown in the first column of Table~\ref{tab:measurements}.

The measurement in the \ttbarjj channel uses a ratio of three-jet to dijet invariant masses ($R_{3/2} = m_{qqb}/m_{qq}$) as its estimator for \mtop. This significantly reduces the sensitivity of the measurement to variations in the JES. A binned minimum-$\chi^2$ fit to the data yields $\mtop = \XZ{173.72}{0.55}{1.01}~\GeV$, where the dominant sources of uncertainty still come from variations in the JES as well as changes to the hadronisation model, as shown in the second column of Table~\ref{tab:measurements}.

The measurement in the \ttbarll channel also uses the invariant mass of the lepton and $b$-jet system (\mlb) as its estimator for \mtop. In this analysis, the total uncertainty in \mtop is minimised by trading statistical precision for decreased systematic uncertainties using a varied requirement on the minimum transverse momentum of the lepton and $b$-jet system (\ptlb). An unbinned maximum-likelihood fit to the data gives $\mtop = \XZ{172.99}{0.41}{0.74}~\GeV$, where the dominant sources of uncertainty come from variations in the JES as well as in the bJES ($b$-jet energy scale), as shown in the third column of Table~\ref{tab:measurements}. With a total uncertainty of 0.84~\GeV, this is the most precise single ATLAS measurement of \mtop. 

\begin{table}[htbp]
\centering
\begin{tabular}{|l|r|r|r|r|}  
\hline
&  single-top  &  all-jets  &  dilepton  &  lepton+jets \\ 
&  [GeV]  &  [GeV]  &  [GeV]  &  [GeV] \\ \hline
\mtop  &  172.2  &  173.72  &  172.99  &  172.08  \\ \hline
Statistical unc.  &  0.7  &  0.55  &  0.41  &  0.39  \\ \hline
\multirow{2}{*}{Dominant syst. unc.} & JES: 1.5 & JES: 0.60 & JES: 0.54 & JES: 0.54 \\
& Had: 0.7 & Had: 0.64 & bJES: 0.30 & $b$-tagging: 0.38 \\ \hline
Total systematic unc. & 2.0 & 1.01 & 0.74 & 0.82 \\ \hline
Total uncertainty & 2.1 & 1.15 & 0.84 & 0.91 \\ \hline
\end{tabular}
\caption{Direct measurements of \mtop in the single-top~\cite{stop}, \ttbarjj~\cite{aj8}, \ttbarll~\cite{dl8}, and \ttbarlj~\cite{lj8} analyses at 8~\TeV, where Had. is hadronisation and (b)JES is the ($b$-)jet energy scale.}
\label{tab:measurements}
\end{table}

\section{New direct \boldmath{\mtop} measurement and combination}

The most recent direct measurement of \mtop at 8~\TeV\ is in the \ttbarlj decay channel. In contrast to the other direct measurements that only used templates of one distribution, this analysis uses three distributions in a 3-D template method that was developed for the \ttbarlj analysis at 7~\TeV~\cite{ljdl7}. The three distributions are the reconstructed top quark mass \mtr, hadronicically decaying $W$ mass \mWr, and ratio of jet transverse momenta
\begin{equation}
    \rbqr  = \frac{p_{\mathrm{T}}^{b_{\rm had}} + p_{\mathrm{T}}^{b_{\rm lep}}}{p_{\mathrm{T}}^{q_1}    + p_{\mathrm{T}}^{q_2}}  
\end{equation}
where $q_1$ and $q_2$ are the light jets assigned to the $W$ boson and $b_{\rm had}$ ($b_{\rm lep}$) is the $b$-jet corresponding to the top quark with the hadronically (leptonically) decaying $W$ boson. 

\begin{figure}[htbp]
\begin{minipage}[b]{.475\linewidth}
\centering
\includegraphics[width=\textwidth]{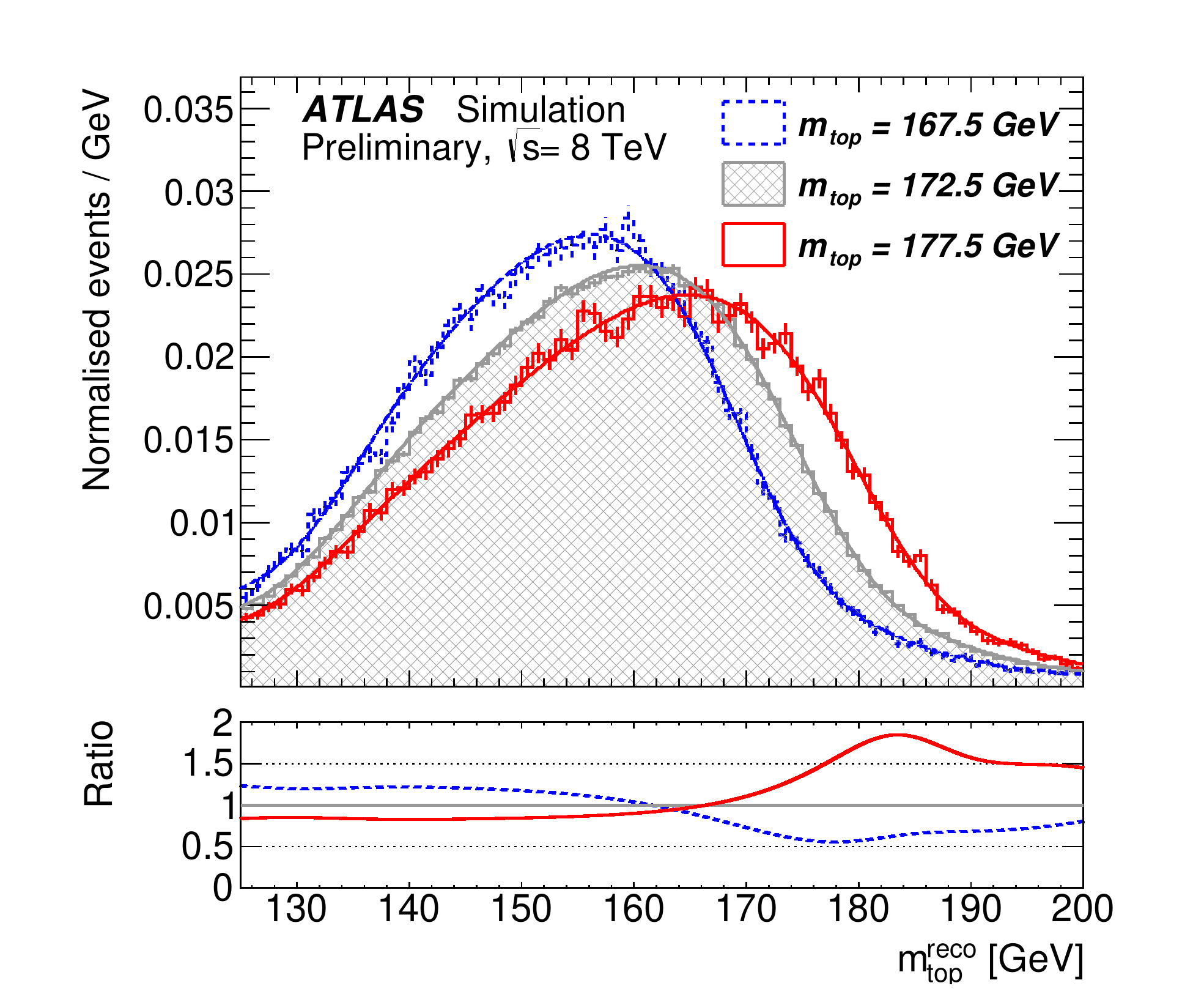}
\subcaption{Template fits to MC samples}\label{fig:temp_a}
\end{minipage}%
\begin{minipage}[b]{.525\linewidth}
\centering
\includegraphics[width=\textwidth]{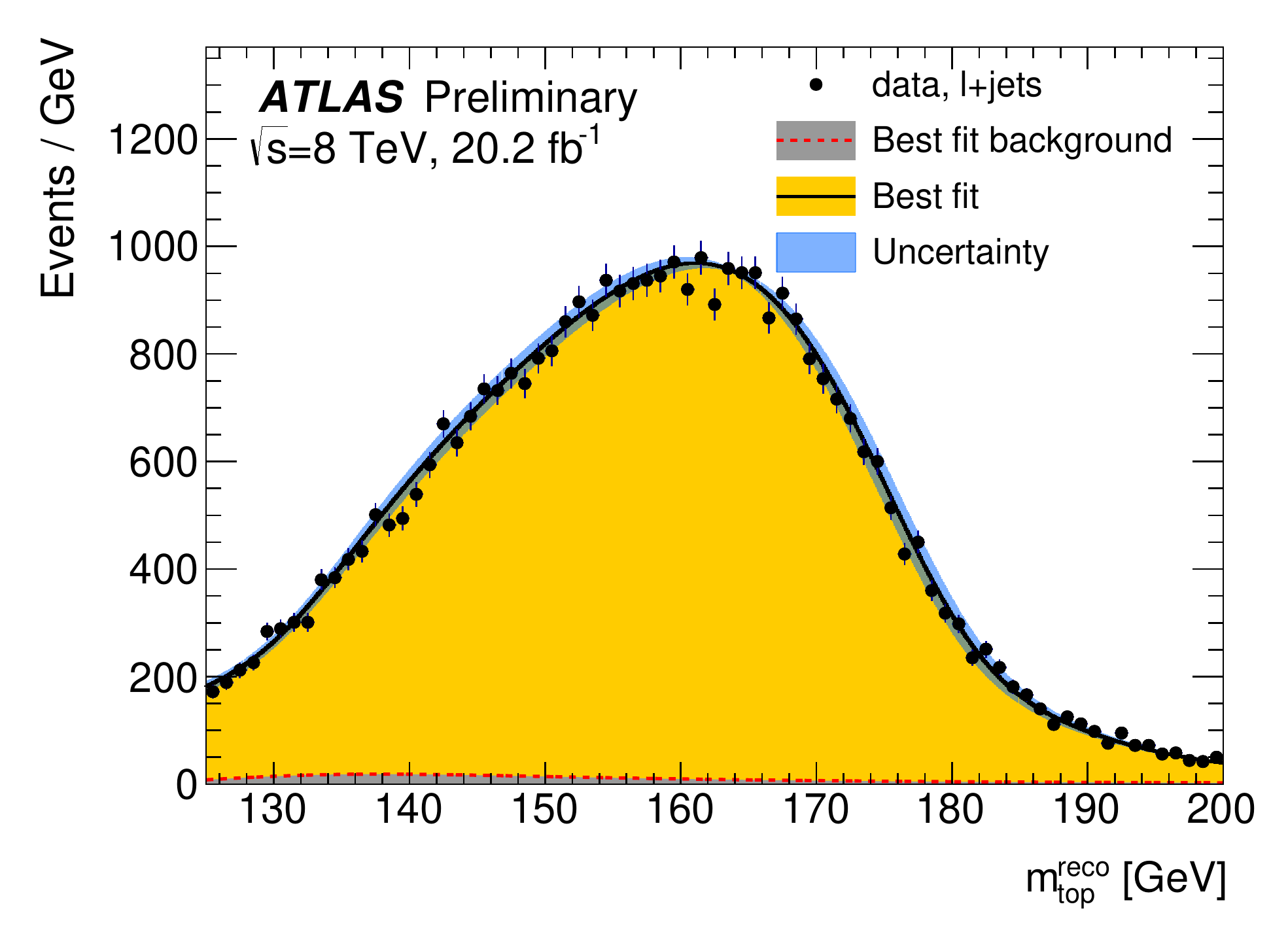}
\subcaption{Likelihood fit to data}\label{fig:temp_b}
\end{minipage}
\caption{(a) Template fits to signal Monte Carlo (MC) events, showing the sensitivity of \mtr to the input \mtop. Each template is overlaid with the corresponding probability density function (pdf). The ratio shown in the lower panel is calculated with respect to the central sample pdf. (b) Results of the likelihood fit to the data in the \mtr distribution. The figure shows the data distribution with statistical uncertainties along with the fitted pdf and its uncertainty.~\cite{lj8}}\label{fig:temp}
\end{figure}

These distributions are fit to analytical functions which are parameterised not only as functions of the input \mtop, but also a jet energy scale factor (JSF) and a $b$-to-light-jet energy scale factor (bJSF), depending on the distribution's sensitivity to changes in all of the above. As an example, the significant sensitivity of the \mtr distribution to changes in the input \mtop is shown in Figure~\ref{fig:temp_a}. The simultaneous measurement of \mtop, JSF, and bJSF reduces the sizeable JES and bJES induced uncertainties in \mtop. 

Similar to the 8~\TeV\ \ttbarll measurement, the total uncertainty in \mtop is minimised, this time with a selection based on a boosted decision tree (BDT) algorithm. The BDT algorithm is used to distinguish events with a correct jet-to-parton matching, with the underlying assumption that these events will have smaller systematic uncertainties. A requirement is made on the BDT output that results in the smallest total uncertainty in \mtop. 

The unbinned 3-D maximum-likelihood fit to the data (shown for the \mtr distribution in Figure~\ref{fig:temp_b}) results in $\mtop = \XZ{172.08}{0.39}{0.82}~\GeV$, where the dominant sources of uncertainty come from variations in the JES and $b$-tagging, as shown in the fourth column of Table~\ref{tab:measurements}. The total uncertainty of 0.91~\GeV\ is a 19\% improvement compared to the same analysis without a BDT requirement, and is altogether a 29\% improvement over the 7~\TeV\ \ttbarlj measurement.

\begin{figure}[htbp]
\begin{minipage}[b]{.5\linewidth}
\centering
\includegraphics[width=\textwidth]{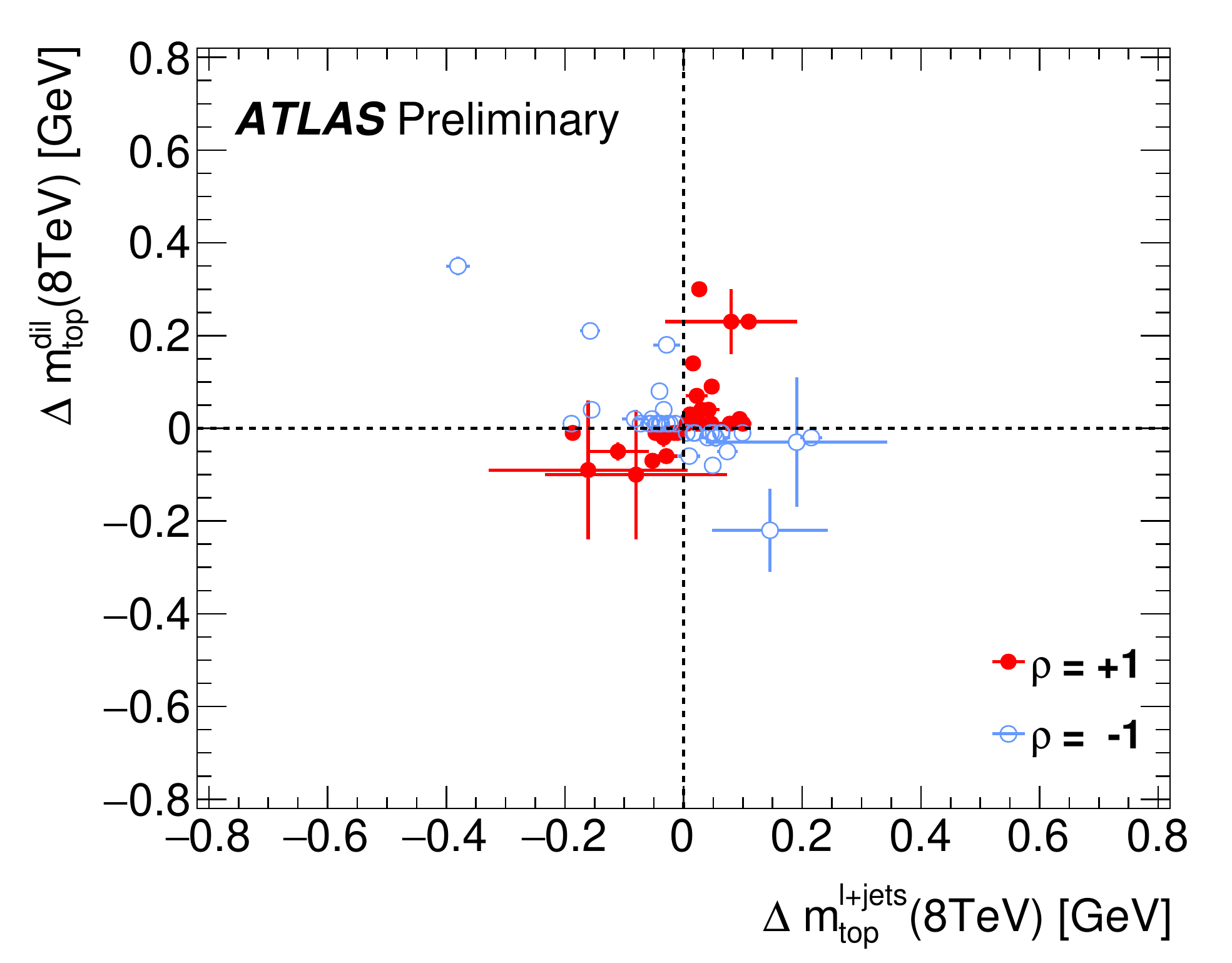}
\subcaption{}\label{fig:cor_a}
\end{minipage}%
\begin{minipage}[b]{.5\linewidth}
\centering
\includegraphics[width=\textwidth]{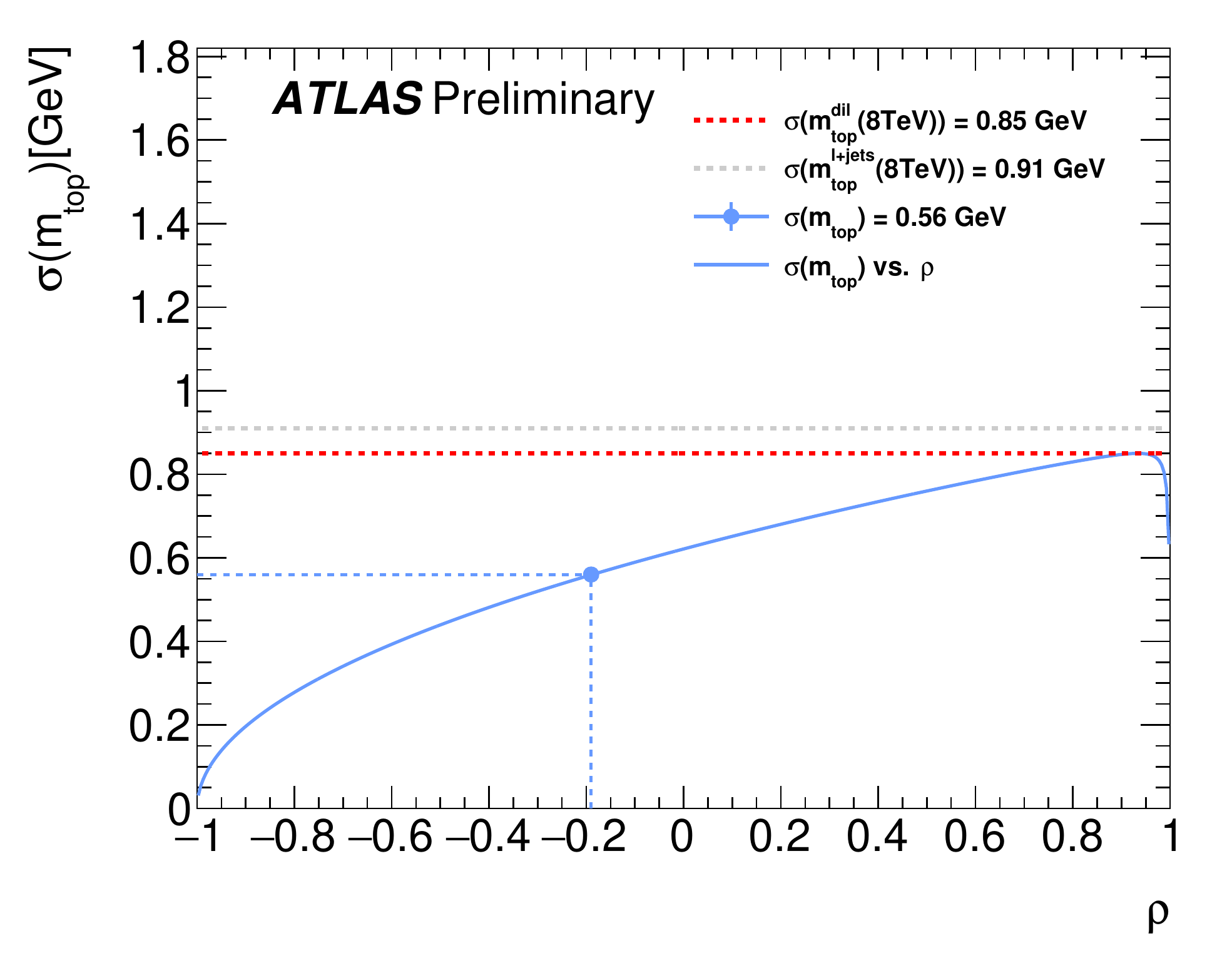}
\subcaption{}\label{fig:cor_b}
\end{minipage}
\caption{(a) The pairwise differences in \mtop when simultaneously varying a pair of measurements for the subcomponents of a systematic uncertainty. (b) The uncertainty of the combination of the two results from 8~\TeV\ data as a function of their correlation (blue full line). The blue point corresponds to the actual correlation.~\cite{lj8}}\label{fig:cor}
\end{figure}

Given the new result, a combination is made of the ATLAS \mtop results in the \ttbarlj and \ttbarll channels at both 7 and 8~\TeV, closely following the strategy of the previous combination of results at 7~\TeV~\cite{ljdl7}. The combination is performed using the best linear unbiased estimate (BLUE) method~\cite{blue1,blue2} for which the central values, the list of uncertainty components,
and the correlations $\rho$ of the estimators for each uncertainty component have to be provided. These correlations are evaluated and can be seen for one pair of measurements in Figure~\ref{fig:cor_a} with the pairwise shift in \mtop ($\Delta{\mtop}$) when simultaneously varying a pair of measurements for each systematic uncertainty.

In Figure~\ref{fig:cor_b}, which shows the uncertainty in the combination of the two most precise results as a function of their correlation, one can easily see the importance of decreasing correlation. This provides for the significant improvement in precision over the most precise single measurement. 

\begin{figure}[htbp]
\begin{minipage}[b]{.5\linewidth}
\centering
\includegraphics[width=\textwidth]{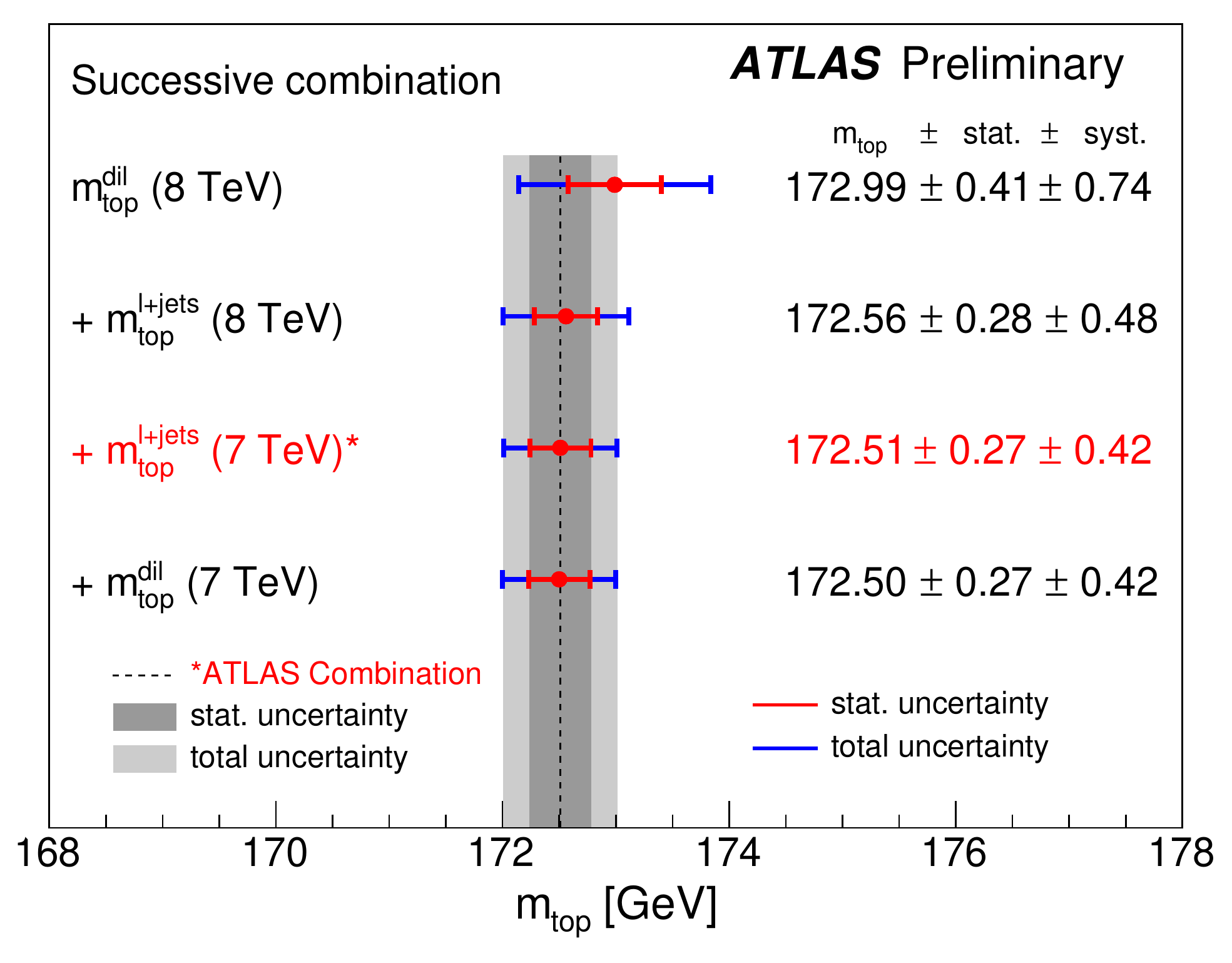}
\subcaption{Significance}\label{fig:combo_a}
\end{minipage}%
\begin{minipage}[b]{.5\linewidth}
\centering
\includegraphics[width=\textwidth]{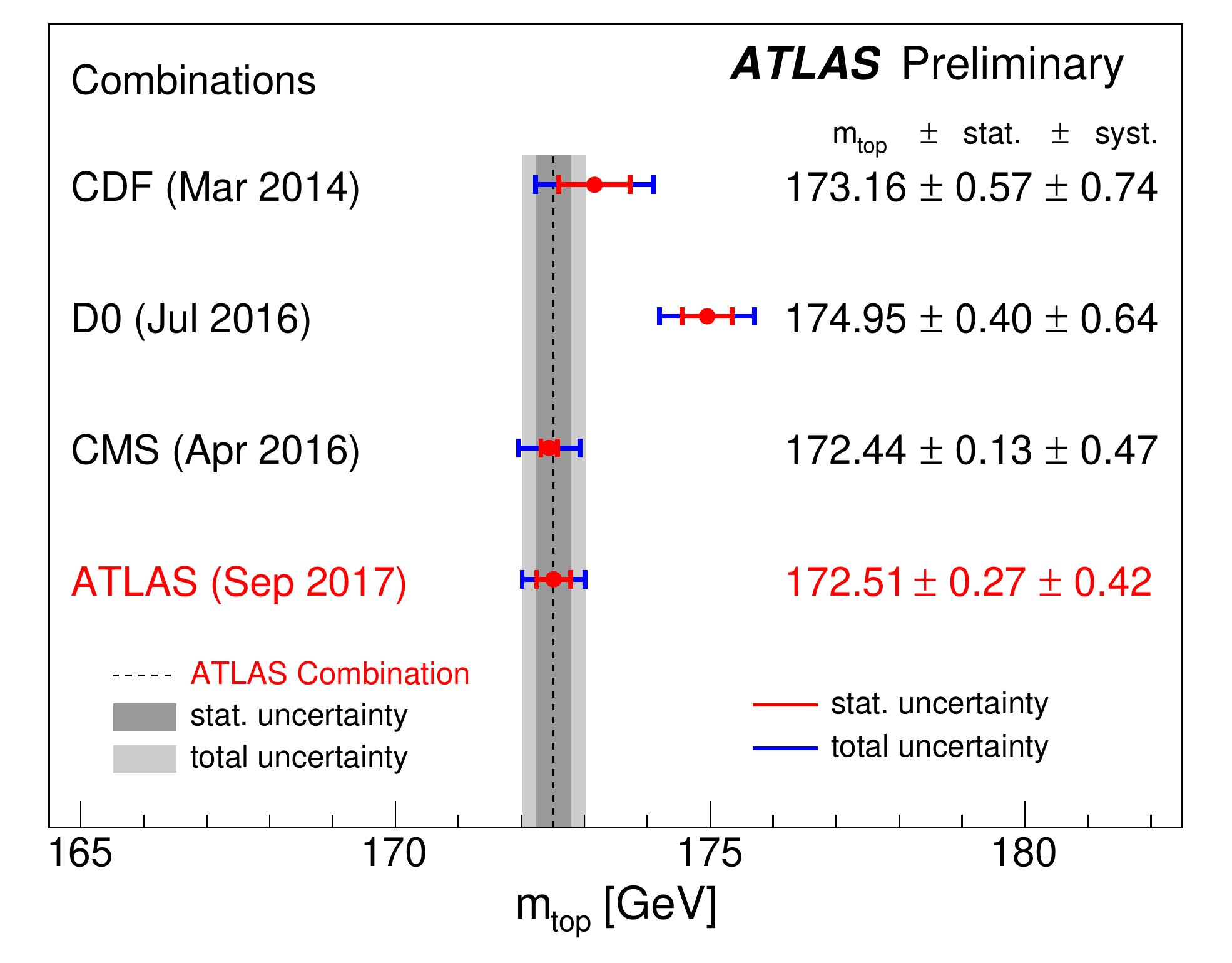}
\subcaption{Comparisons}\label{fig:combo_b}
\end{minipage}
\caption{Results of the combination~\cite{lj8}. Figure (a) shows the combined result when successively adding results to the most precise one. Each line of this figure shows the combined result when adding the result indicated by a '+'  to the combination. The new ATLAS combination is shown in red and indicated by the star. Figure (b) shows the combined result for \mtop per experiment from the latest combinations performed by the individual experiments. In both figures the vertical band corresponds to the new ATLAS combination of \mtop.}\label{fig:combo}
\end{figure}

Figure~\ref{fig:combo_a} shows the combination of the four measurements with one addition at a time, in order of their significance in BLUE. The red star indicates the new ATLAS combined measurement of \mtop as adding the \ttbarll measurement at 7~\TeV results in an insignificant change compared to the statistical precision of the uncertainty in the previous combined result. This distinguishing power is available thanks to the individual measurements having evaluated the statistical precision of the systematic uncertainties. The total uncertainty in the combined result quoted is known to $\pm 0.4~\GeV$.

The new ATLAS combined result of $\mtop = \XZ{\CombVal}{\CombSta}{\CombSys}~\GeV$ is shown in relation to other combined results in Figure~\ref{fig:combo_b}. The result from CMS~\cite{cms} is consistent with that of ATLAS and has a similar total uncertainty. The CDF result~\cite{cdf} is consistent with both LHC results, while that of D0~\cite{d0} is somewhat larger. 

\section{Conclusion}

ATLAS has made several measurements of the top quark mass using LHC proton-proton collisions at a 
CM energy of 8~\TeV. These include an indirect determination of the top quark pole mass, performed using lepton differential cross-sections, and direct measurements of the top quark mass in each \ttbar decay channel as well as in the $t$-channel of single-top-quark production, all performed using template methods. With the completion of the new preliminary direct measurement in the \ttbarlj channel, a new preliminary ATLAS combination was also performed, resulting in $\mtop = \XZ{\CombVal}{\CombSta}{\CombSys}~\GeV$ with a total uncertainty of $\XZtot{\CombUnc}{\CombUncStab}~\GeV$, where the quoted uncertainty in the total uncertainty is statistical. The pole mass of $\mtpole = 173.2 \pm 0.9\,\mathrm{(stat)} \pm 0.8\,\mathrm{(syst)} \pm 1.2\,\mathrm{(theo)}\,\GeV$ is consistent with the direct measurement combination within uncertainties.

Two crucial general strategies were used to reach the 0.29\% relative precision in the combined \mtop. First, the individual analyses, being systematically limited, traded statistical for systematic precision to achieve reduced total uncertainties. Second, care was taken to minimise, as well as properly evaluate, the correlations between the individual measurements.


\end{document}